# Topological States in Dimerized Quantum-Dot Chains Created by Atom Manipulation

Van Dong Pham[1,*], Yi Pan[1,2], Steven C. Erwin[3], Felix von Oppen[4], Kiyoshi Kanisawa[5], and Stefan Fölsch[1,†]

[1]*Paul-Drude-Institut für Festkörperelektronik, Hausvogteiplatz 5-7, Leibniz-Institut im Forschungsverbund Berlin e. V., 10117 Berlin, Germany*

[2] *Center for Spintronics and Quantum Systems, State Key Laboratory for Mechanical Behavior of Materials, Xi'an Jiaotong University, Xi'an 710049, China*

[3] *Center for Computational Materials Science, Naval Research Laboratory, Washington, DC 20375, USA*

[4]*Dahlem Center for Complex Quantum Systems and Fachbereich Physik, Freie Universität Berlin, 14195 Berlin, Germany*

[5]*NTT Basic Research Laboratories, NTT Corporation, 3-1 Morinosato Wakamiya, Atsugi, Kanagawa, 243-0198, Japan*

Topological electronic phases exist in a variety of naturally occurring materials but can also be created artificially. We used a cryogenic scanning tunneling microscope to create dimerized chains of identical quantum dots on a semiconductor surface and to demonstrate that these chains give rise to one-dimensional topological phases. The dots were assembled from charged adatoms, creating a confining potential with single-atom precision acting on electrons in surface states of the semiconductor. Quantum coupling between the dots leads to electronic states localized at the ends of the chains, as well as at deliberately created internal domain walls, in agreement with the predictions of the Su-Schrieffer-Heeger model. Scanning tunneling spectroscopy also reveals deviations from this well-established model manifested in an asymmetric level spectrum and energy shifts of the boundary states. The deviations arise because the dots are charged and hence lead to an onsite potential that varies along the chain. We show that this variation can be mitigated by electrostatic gating using auxiliary charged adatoms, enabling fine-tuning of the boundary states and control of their quantum superposition. The experimental data, which are complemented by theoretical modeling of the potential and the resulting eigenstates, reveal the important role of electrostatics in these engineered quantum structures.

[*] pham@pdi-berlin.de
[†] stefan.foelsch@pdi-berlin.de

# I. INTRODUCTON

The boundary states of topological insulators exist within the bulk energy gap and cannot be removed by perturbations that leave the bulk spectrum gapped [1,2]. This robustness may enable new applications in information storage and processing [3,4,5]. An attractive way of reaching this goal is the experimental realization of topological electronic phases in engineered condensed-matter systems. This has already been done in a few nanoscale materials such as graphene nanoribbons [6,7], polymer chains created by on-surface synthesis [8], and artificial lattices created with a scanning probe tip on a metal surface [9,10,11,12]. Boundary states with tunable topological character were demonstrated in these systems, marking an important step towards their use in future technologies.

In this work we created dimerized chains of quantum dots on a semiconductor surface using atom manipulation by cryogenic scanning tunneling microscopy (STM) [13]. We then used scanning tunneling spectroscopy measurements to demonstrate the existence of boundary states associated with the chain ends and with domain walls deliberately fabricated in their interior. The quantum coupling between the dots leads to molecular-like states having sub-ångström precision conferred by the atom-by-atom method used to create the chains. By varying the alternating electron hopping between the dots we demonstrated that the resulting linear combinations of dot states comprising the boundary states are consistent with the Su-Schrieffer-Heeger (SSH) model of one-dimensional (1D) topological phases [14,15,16]. We therefore conclude that the observed boundary states are topological in origin, reflecting the topological nature of the bulk phase via the bulk-boundary correspondence.

We also observed deviations from the SSH model, manifested as an asymmetry in the energy level spectra. This deviation arises because the dots are charged and create an onsite potential that breaks the sublattice symmetry. The symmetry-breaking shifts the boundary states in energy, alters their wave functions, and modifies their quantum superposition in finite chains. We demonstrate that these effects can be tuned by electrostatic gating using auxiliary charged adatoms, opening the door to manipulating the boundary states. The experimental data are complemented by theoretical modeling of the onsite potential due to the adatoms [17]; the corresponding eigenstates calculated within a tight-binding model reveal the crucial role of electrostatics in these engineered quantum structures. Our results demonstrate that the atom-by-atom assembly in combination with the screening properties of the semiconductor environment provides a unique degree of spatial control over the confining potential and thereby the resulting electronic states.

## II. RESULTS AND DISCUSSION

### A. Single quantum dots

We assembled quantum dots, each consisting of six In atoms, on an InAs(111)A surface grown by molecular beam epitaxy (details on the growth and sample preparation are given in Appendix A). This surface offers two essential ingredients for the fabrication of atomic-scale quantum dots, namely an intrinsic surface state [18] and the presence of native In adatoms (concentration roughly 0.005 monolayers) that can be repositioned by the STM tip with sub-ångström precision [18,19,20,21]. The In adatoms are positively charged and hence assemblies of these adatoms confine surface-state electrons [18]. InAs(111)A exhibits a 2×2 In vacancy reconstruction with a hexagonal surface unit cell and a lattice constant (vacancy spacing) of $a'$=8.57 Å. The reconstruction renders the surface chemically inert [22] and the In vacancies define the allowed sites of the native In adatoms. A linear arrangement of six In adatoms in adjacent vacancy sites creates a quantum dot or "artificial atom" that gives rise to a confined state at ~0.1 eV below the Fermi level [18].

**B. Quantum-dot dimers**

We next created two identical quantum dots separated by spacings 3, 4, and 5 in units of √3a′ [Fig. 1(a)] and probed their electronic states using scanning tunneling spectroscopy measurements of the differential tunnel conductance $dI/dV$. By using a lock-in technique and peak-to-peak modulation of 5 mV, we obtain an energy resolution [23] of $\Delta E$=4.7 meV at the present measurement temperature of 5 K. Conductance spectra from these quantum-dot "molecules" reveal a bonding (σ) and an antibonding state (σ*) [18,21], as shown in the lower panel of Fig. 1(a). The spectra cover the bias range from zero to −0.25 V, equivalent to an energy range from the Fermi level of the InAs surface to 0.25 eV below. The bonding-antibonding splitting $\Delta_{\sigma-\sigma*}$, which reflects the degree of quantum coupling between the two dots, decreases as their spacing increases. This splitting allows us to define an electron hopping amplitude $t = e\Delta_{\sigma-\sigma*}/2$ = 55, 39, and 29 meV for spacings of 3, 4, and 5 units, respectively (hopping amplitudes are defined as positive quantities throughout this paper).

The width of the conductance peaks is 10 to 15 mV, smaller by a factor of 5-10 than $\Delta_{\sigma-\sigma*}$. This makes it possible to probe individual quantum states for more extended quantum-dot assemblies, as discussed below. The energy resolution of 4.7 meV implies that the observed peak width is primarily limited by lifetime broadening resulting from the coupling of the confined electrons to the semiconductor environment. In contrast, on metal surfaces, the effect of lifetime broadening is generally larger [24,25].

In addition to the bonding-antibonding splitting, the spectra in Fig. 1(a) reveal that the σ–σ* doublet is downshifted from the energy of an individual $In_6$ dot [gray curves in Fig. 1(a)]. The shift $\Delta_s$ decreases as the interdot separation increases. We interpret the shift as electrostatic and arising from the potential change that each dot experiences from the other.

**C. Quantum-dot-dimer chains**

We next created chains of identical quantum dots with various spacings between the dots. The STM topography images in Figs. 1 (b-d) illustrate some of the chains investigated. The top panel shows a chain of $N = 10$ dots with dimerized (alternating) spacing of 4 and 3 units, referred to here as (4,3) chains. This arrangement can be described as a sequence of five unit cells each containing two dots, one on sublattice $A$ and the other on sublattice $B$. The dimerized spacing was chosen so that the intracell hopping $t_1 = 39$ meV is smaller than the intercell hopping $t_2 = 55$ meV, and hence the weaker hopping occurs at the ends, $t_1/t_2 \cong 0.7$. In the tight-binding description of the SSH model (outlined in Appendix B) this choice of alternating hopping between degenerate atomic orbitals leads to the emergence of a localized state at both ends of a finite chain, one residing on sublattice $A$ and the other on sublattice $B$ [26]. In the limit of a long chain, these end states are located in the energy band gap of the dimerized chain and decay exponentially into the bulk. The overlap of the end states leads to an energy splitting between their odd and even superpositions.

We searched for end states in dimer chains with the weaker hopping occurring at the ends, $t_1/t_2<1$. A (4,3) chain of this kind consisting of $N=8$ dots is shown in the top panel of Fig. 2(b). The colored circles indicate STM tip positions at which the conductance spectra displayed in Fig. 2(a) were recorded. A sequence of conductance peaks is observed which can be associated with eight states labelled by the quantum number $n$ plus an extra peak denoted $P$ reflecting a state at higher energy, about 70 meV below the Fermi level. The lower-lying eight peaks (marked by vertical bars) correspond to "molecular" states of the chain. We focus on the states $n = 4,5$ situated in the gap between the lower ($n = 1,2,3$) and upper states ($n = 6,7,8$). Their spatial distribution along the axis of the chain can be visualized by the conductance probed at constant tip height at fixed sample bias. Performing this scan as a function of bias yields the conductance map in the lower panel of Fig. 2(b). The map reveals that the states with $n=4$ and 5 have maximum probability density at the end dots and reside predominantly on sublattice $A$ ($B$) sites in the left (right) half of the chain. In contrast, the other states have probability density concentrated in the chain's interior; this is especially clear for the lower set of states. We thus identify the states $n = 4$ and 5 as the odd (labelled $e_1$) and even superposition ($e_2$) of the chain end states, respectively.

These chains exhibit topological end states because the hopping $t_1$ at the ends is weaker and hence $t_1/t_2<1$. The reverse situation $t_1/t_2>1$ realizes a topologically trivial chain without end states. We confirmed this different behavior in short chains $N = 6$ by measuring their normalized conductance spectra [Fig. 3]. (The normalization $(dI/dV)/(I/V)$ is useful here because it yields an approximate measure of the density of states at the surface [27], allowing one to compare peak magnitudes on similar footing.) We address the case of weaker hopping at the ends in more detail, by comparing (5,3) chains (alternating spacing of 5 and 3 units, $t_1/t_2 \cong 0.5$) and the (4,3) chains ($t_1/t_2 \cong 0.7$) discussed above. In general, the localization of the end states $e_1$ and $e_2$ at the outer dots becomes more pronounced as the ratio $t_1/t_2$ is reduced. Conversely, as $t_1/t_2$ is made larger, the resulting delocalization will increase the overlap of $e_1$ and $e_2$ and thereby their energy splitting. Indeed, as $t_1/t_2$ is made larger by going from (5,3) to (4,3) chains, the conductance spectra show that the $e_1$–$e_2$ splitting increases from 12 mV to 25 mV. In addition, the probability density at the inner dots (dot indexes $i$=2 to 5) increases at the expense of the outer dots ($i$=1 and 6). Finally, the spectra reveal that the energy range of the molecular states is smaller for (5,3) than for (4,3) chains, in agreement with the fact that the total energy bandwidth scales with $t_1$+$t_2$ for an atomic chain with alternating hopping [14,16].

We turn to the case of stronger hopping at the ends, using the example of a (3,4) chain (alternating spacing of 3 and 4 units, $t_1/t_2 \cong 1.4$). The spectra reveal six molecular states but the energy spacing $E_4$–$E_3$ between states $n$ = 3 and 4 is now 60 meV, several times larger than the end-state splitting observed before. In addition, the states $n$=3 and 4 have their probability density maxima at inner dots and thus lack localization at the ends. Figure 3(b) is a plot of the normalized energy spacing $(E_4 - E_3)/t_2$ versus the hopping ratio $t_1/t_2$ for these $N$ = 6 chains and a (4,4) chain with $t_1$=$t_2$ [28]. End states exist when $t_1/t_2$<1 while they are absent when $t_1/t_2 \geq 1$. Also shown (black curve) is the tight-binding result for a dimerized chain with $N$=6 assuming equal site energies (cf. Appendix B). Our data are in excellent agreement with this phase diagram showing that $t_1/t_2$<1 leads to localized end states while $t_1/t_2 \geq 1$ does not.

The existence of topological end states in the SSH model is closely linked to the symmetries of the model. In particular, the existence of midgap end states requires sublattice or chiral symmetry. This symmetry implies equal site energies of the $A$ and $B$ sublattices as well as the absence of next-nearest-neighbor hopping and leads to an energy level spectrum that is symmetric about the center of the gap (cf. Appendix B). It is obvious from the conductance spectra in Figs. 2 and 3 that the measured energy levels $E_n$ do not show this symmetry. The cause of the observed asymmetry is readily understood: the dots are charged (because the In adatoms themselves are charged) and hence create a varying onsite potential along the chain that is higher at the ends than in the middle, breaking the sublattice symmetry. The qualitative behavior of such a potential is to increase energy spacings in the lower half of the spectrum and compress them in the upper half, in agreement with Figs. 2 and 3. We turn next to analyzing and modeling this behavior quantitatively.

To make a systematic analysis of the energy level structure we investigated chains assembled at different surface locations, on different samples, and during different experimental runs. Figure 4(a) summarizes the energies measured for 17 (4,3) chains with $N = 4,6,8,10$. The scatter in the data reflects electrostatic potential disorder due to residual charged defects [29,30,31]. For the MBE-grown samples used in our study these effects are small yielding a variation in relative conductance peak positions and level spacings with a standard deviation of ~1 mV [18]. The absolute values have a standard deviation of ~5 mV [28]. The entire data set reveals that the level spectrum at given $N$ is strongly asymmetric about the midpoint of $e_1$ and $e_2$. In addition, with increasing $N$ there is a significant downward shift of the entire level spectrum.

These trends are well reproduced by the theoretical energies indicated by large empty circles in Fig. 4(a). We used a nearest-neighbor tight-binding Hamiltonian with experimental hoppings $t_1$, $t_2 = 39, 55$ meV and onsite energies $e\langle V_i \rangle$ defined by the expectation value $\langle V_i \rangle = \langle \varphi_i | V | \varphi_i \rangle$ where $\varphi$ is a hydrogenic orbital located at each dot center and $V$ is the potential from all the In adatoms screened by the 2D electron gas at the InAs surface [17,32,33,34,35] (see Appendix C for details). We assumed each In adatom is positively charged and incompletely ionized with $q < 1$ when assembled into a chain. A least-squares fit to the experimental energies yields $q = 0.69$, which is in reasonable agreement with an estimate obtained from minimizing the energy gained by the adatoms from ionization plus the energy cost from their Coulomb interaction, as discussed in Appendix D.

We turn briefly to the extra spectral peak denoted $P$ in Figs. 2 and 3 and show that it is of different origin than the molecular states. Conductance spectra of $N = 4,6,8,10$ chains show that this peak occurs in the same energy range for all $N$ and is significantly larger than the conductance peaks associated with the molecular states [Fig. 4(b)]. We propose that the peak *P* arises from the bound state [36] of the 2D electron gas on InAs(111)A [32,33,34,35]. This electron accumulation layer arises from defects (such as the native In adatoms) that create donor states above the conduction band minimum and thereby a downward band bending near the surface. While the accumulation layer extends over a region ~20 nm below the surface, it is still sensitive to the scattering at surface defects as revealed previously by the STM observation of quasi-particle interference patterns [34]. This suggests that the accumulation layer undergoes lateral confinement in the presence of quantum-dot dimer chains as well.

### D. Chains with internal domain walls

The ends of a finite dimerized chain are part of the broader class of domain walls that break a sublattice symmetry. To further explore this class we introduced actual domain walls interrupting the sublattice order within a chain. Within the SSH model this type of defect creates a midgap state [26]. The topography image in Fig. 5(a) shows an $N = 10$ chain with a "light" domain wall centered at site $i=5$. This defect means that the rightmost dot ($i=10$) has weaker coupling. Therefore, the boundary state of the internal domain wall will overlap with that of the right end state, leading to even and odd superpositions. For $N = 10$ these correspond to the states $n = 5$ and 6 in the middle of the spectrum. In the absence of any onsite potential their wave-function amplitudes along the chain would be identical, differing only in signs. In the presence of the onsite potential, however, these two superpositions are no longer degenerate. Instead, they experience an energy shift that depends on the position of the boundary. The domain-wall state is lower in energy than the end state because the potential is lower in the interior. Hence the lower-lying superposition ($n=5$) has more domain-wall character and the higher-lying superposition ($n=6$) has more end-state character.

This effect is easily seen in the conductance maps for $n=5$ and $n=6$ [Fig. 5(a) upper panel]. These maps show good overall agreement with the squared wave-function amplitudes $|\Psi_5|^2$ and $|\Psi_6|^2$ obtained from our tight-binding model [Fig. 5(a) lower panel].

The ten dots can also be rearranged to form a "heavy" domain wall, here centered at site $i=6$, plus a weakly coupled dot at the right end [Fig. 5(b)]. The conductance maps again reveal the probability density for the states $n=5$ and $n=6$ to be very different: the lower-lying superposition (having domain-wall character) shows probability density predominantly on the dots $i=5$ and 7 adjacent to the domain wall position. This is readily understood within the fully dimerized limit of a SSH chain in which the heavy domain wall corresponds to a discrete trimer having a zero-energy state with wave-function amplitude exclusively on its outer sites. This typical signature of a heavy-wall boundary state is also evident in the theoretical amplitudes $|\Psi_5|^2$ and $|\Psi_6|^2$ [Fig. 5(b) lower panel].

**E. Fine-tuning by electrostatic gating**

Finally, we show that boundary states can be manipulated by adding auxiliary charged adatoms to create localized electrostatic gating. We used the ten-dot dimer chain discussed in Fig. 5(a) and positioned four additional adatoms near its ends as shown in Fig. 6. These gating atoms flatten the onsite potential along the chain [37] and thereby bring the chain closer to a SSH model with constant onsite energies. This flattening partially restores the degeneracy of the two boundary states which was absent in the ungated chain: conductance maps (Fig. 6 center panel) show that the probability density distributions of the even ($n=5$) and odd superpositions ($n=6$) are now very similar. The same effect is also evident in the theoretical squared wave-function amplitudes in the lower panel of Fig. 6(a), which were obtained by explicitly including the electrostatic effect of the auxiliary adatoms. These results demonstrate that boundary states can be tuned by modifying the electrostatic potential around the dimer chain.

## III. CONCLUSIONS

We have shown that boundary states in dimerized chains of identical quantum dots have the properties predicted by the SSH model of 1D topological phases. Deviations from this model – asymmetry in the energy level spectrum and shifts of the boundary state energies – arise from a varying onsite potential created by the partial ionization of the individual adatoms comprising the dots. We have shown that auxiliary adatoms can be used to create electrostatic gates that flatten this potential and largely restore sublattice symmetry, thus revealing the important role of electrostatics in these engineered quantum structures. The approach to simulate electronic structure by atom manipulation on a semiconductor surface, as we have demonstrated it here, is general and versatile. It can be readily extended to quantum dot arrays forming topological quantum gates [38] and 2D-coupled architectures [39] based on the SSH model or to other 2D artificial lattices giving rise to topological or strongly correlated electronic states [40,41].

## APPENDIX A: SAMPLE GROWTH AND PREPARATION

20-nm-thick undoped InAs layers were grown by molecular beam epitaxy (MBE) on an InAs(111)A substrate (purchased from *Wafer Technology Ltd*) to prepare the In-terminated InAs(111)A surface with its intrinsic (2×2) In-vacancy reconstruction as monitored *in situ* by reflection high energy electron diffraction. Directly after the MBE growth, the surface was capped by an amorphous layer of arsenic and transferred under ambient conditions to the ultra-high vacuum chamber of the STM apparatus. The As capping layer was then desorbed by annealing at 630 K and the sample was loaded into the microscope. InAs(111)A samples prepared in this way showed the same surface features as MBE-grown and *in situ* investigated samples.

**APPENDIX B: TIGHT-BINDING DESCRIPTION**

The SSH model was originally conceived to describe charge-carrier transport in conducting polymers [14,15,16]. For fixed lattice distortions, one obtains a dimerized electronic tight-binding model in which electrons hop with alternating amplitudes between nearest-neighbor sites. In second quantization, the Hamiltonian takes the form

$$\mathcal{H} = -\sum_j [t + (-1)^j \Delta/2] c_j^\dagger c_{j+1} + h.c. \tag{A1}$$

where $c_j$ annihilates a fermion at site $j$. The dimerization $\Delta$ modulates the nearest-neighbor hopping along the chain, so that hopping alternates between $t_1=t-\Delta/2$ and $t_2=t+\Delta/2$. For infinite chains the resulting band structure,

$$E(k) = \pm\sqrt{2t^2[1+\cos(ka)] + \Delta^2/2\,[1-\cos(ka)]}, \tag{A2}$$

is gapped for both positive and negative $\Delta$, with a gap closing for $\Delta=0$ separating two topologically distinct phases ($k$ is the wave vector and $a$ the lattice constant).

Finite chains comprising $N$ sites exhibit midgap end states when the outermost bonds exhibit weaker hopping. No such end states exist when the outermost bonds are strong. The midgap nature of the end states is protected by sublattice (sometimes referred to as chiral) symmetry. In the experiment, chiral symmetry is explicitly broken by a spatially varying onsite energy $\epsilon_j$,

$$\mathcal{H} = \sum_j \{-[t + (-1)^j \Delta/2] c_j^\dagger c_{j+1} + h.c. + \epsilon_j c_j^\dagger c_j\}. \tag{A3}$$

Chiral symmetry would also be broken by next-nearest-neighbor hopping, but we find that this is negligible in our experimental system (see below). When chiral symmetry is broken, the end states can shift away from the gap center (weakly broken chiral symmetry) or even merge into the bands (strongly broken chiral symmetry). However, two topologically distinct phases can still be defined in terms of the presence or absence of half-integer polarization charges at the ends as long as the system remains bulk inversion symmetric.

The end states are a consequence of the bulk-boundary correspondence, which also predicts midgap states localized at domain walls at which $\Delta$ changes sign. A heavy domain wall has two neighboring strong bonds, a light domain wall two neighboring weak bonds.

In first quantization, the SSH model becomes a matrix Hamiltonian. For example, for $N=6$ sites the Hamiltonian takes the form

$$H = \begin{pmatrix} 0 & t_1 & 0 & 0 & 0 & 0 \\ t_1 & 0 & t_2 & 0 & 0 & 0 \\ 0 & t_2 & 0 & t_1 & 0 & 0 \\ 0 & 0 & t_1 & 0 & t_2 & 0 \\ 0 & 0 & 0 & t_2 & 0 & t_1 \\ 0 & 0 & 0 & 0 & t_1 & 0 \end{pmatrix}. \tag{A4}$$

For identical (degenerate) atomic orbitals, all site energies (diagonal entries) can be chosen as zero. The resulting energy level spacing $(E_4 - E_3)/t_2$ analyzed in Fig. 3(b) is depicted as a black curve in that diagram.

To model the experimentally observed energy level spectra, we extended the tight-binding description and chose nonzero onsite energies to account for the variation in onsite potential, as outlined in Appendix C. Spatially varying onsite energies shift the end states away from the center of the gap and introduce asymmetries between the bands. Spectral asymmetry can also result from hoppings between next-nearest neighbors. However, next-nearest-neighbor hopping alone does not reproduce the overall downward shift of the levels that is observed as $N$ increases [cf. Fig. 4(a)]. We also find that any nonzero value of these additional hopping terms lowers the quality of the least-squares fit to the experimental energies. It is therefore concluded that next-nearest-neighbor hopping can be safely neglected in the experimental dimer chains.

**APPENDIX C: THEORETICAL ELECTROSTATIC POTENTIAL**

The electrostatic potential was calculated based on the result derived in Appendix B of Ref. [17] for the potential $V(r)$ of a point charge $q$ screened by a two-dimensional electron gas,

$$V(r) = (q/\kappa) \int_0^\infty k(k+s)^{-1} J_0(kr) e^{-kd} dk. \tag{A5}$$

Here, $\kappa$ is the average of the InAs bulk static dielectric constant (15.15) and that of vacuum, $s$ is a screening constant written as $s = 2m^* e^2/\hbar^2 \kappa$, $J_0$ is the Bessel function of order zero, and $d$ is the distance (here set to 1 Å) from the point charge to the measurement plane. We used the effective mass $m^* = 0.05 m_e$, consistent with the experimental value observed for the 2D electron gas at the InAs(111)A surface [34]. From the potential $V$ for a chain of $N$ quantum dots we calculated the expectation value $\langle V_i \rangle = \langle \varphi_i | V | \varphi_i \rangle$ to define onsite energies $e\langle V_i \rangle$ for each site $i=1$ to $N$ in the tight-binding Hamiltonian. For simplicity, we assumed $\varphi$ to be a hydrogenic orbital with a Bohr radius of $a_0=17.5$ Å as obtained from the spatial conductance contour measured for the confined state of an individual In$_6$ dot.

**APPENDIX D: CHARGE STATE OF INDIUM ATOMS IN A CHAIN OF DOTS**

We consider a chain of $N$ dots each consisting of $M$ indium adatoms. An isolated adatom is known to be fully ionized ($q = +1$) but when these are assembled into a chain, the resulting cost from their mutual Coulomb interaction competes with the energy gained by ionization of each adatom. As a result, the ionization of the indium atoms in a chain of dots may be incomplete and hence the average charge state $q$ of each atom will be less than 1. Here we evaluate both energy terms and minimize their sum to obtain a rough quantitative estimate for $q(N,M)$.

An isolated adatom has charge +1 because an electron is transferred from an adatom state to a surface state with lower energy, thereby lowering the energy of the system by $E_0$. We estimate $E_0$ as the difference between the energy of unoccupied In$_{ad}$ states (at about 0.7 eV) and the Fermi level. Thus a chain of $NM$ adatoms, each with charge $q$, gains an energy $qNME_0$.

The magnitude of the Coulomb energy cost depends on the specific arrangement of the atoms. Here we used (4,3) chains; the results for (5,3) chains are very similar. We computed the total electrostatic energy assuming each atom has charge $q$ and creates a 2DEG-screened potential $V(\mathbf{r})$ exactly as assumed by the tight-binding modeling discussed in subsection C of the main text. The resulting total energy is quadratic in $q$ and can be easily minimized numerically for each value of $N$ and $M$ to obtain an estimate for $q(N,M)$ as shown in Fig. 7. These results should be viewed as very rough estimates. For the typical chains discussed in Fig. 2 of $N = 8$ dots, each with $M = 6$ atoms, the predicted $q$ is 0.85.

**ACKNOWLEDGEMENTS**

VDP, YP, and SF acknowledge financial support by the Deutsche Forschungsgemeinschaft (FO362/4-2, FO362/5-1). SCE was supported by the Office of Naval Research through the Naval Research Laboratory's Basic Research Program. FvO was supported by CRC 183 of Deutsche Forschungsgemeinschaft.

**References**

[1] M. Z. Hasan and C. L. Kane, *Topological Insulators*, Rev. Mod. Phys. **82**, 3045 (2010).

[2] F. D. M. Haldane, Nobel Lecture: *Topological Quantum Matter*, Rev. Mod. Phys. **89**, 040502 (2017).

[3] S. Ornes, *Topological Insulators Promise Computing Advances, Insights into Matter Itself*, Proc. Natl Acad. Sci. USA **113**, 10223 (2016); and references therein.

[4] M. He, H. Sun, and Q. L. He, *Topological Insulator: Spintronics and Quantum Computations*, Front. Phys. **14**, 43401 (2019); and references therein.

[5] M. J. Gilbert, *Topological electronics*, Commun. Phys. **4**, 70 (2021); and references therein.

[6] D. J. Rizzo, G. Veber, T. Cao, C. Bronner, T. Chen, F. Zhao, H. Rodriguez, S. G. Louie, M. F. Crommie, and F. R. Fischer, *Topological Band Engineering of Graphene Nanoribbons*, Nature (London) **560**, 204 (2018).

[7] O. Gröning, S. Wang, X. Yao, C. A. Pignedoli, G. Borin Barin, C. Daniels, A. Cupo, V. Meunier, X. Feng, A. Narita, K. Müllen, P. Ruffieux, and R. Fasel, *Engineering of Robust Topological Quantum Phases in Graphene Nanoribbons*, Nature (London) **560**, 209 (2018).

[8] B. Cirera, A. Sánchez-Grande, B. de la Torre, J. Santos, S. Edalatmanesh, E. Rodríguez-Sánchez, K. Lauwaet, B. Mallada, R. Zbořil, R. Miranda, O. Gröning, P. Jelínek, N. Martín, and D. Ecija, *Tailoring topological order and π-conjugation to engineer quasi-metallic polymers*, Nat. Nanotechnol. **15**, 437 (2020).

[9] R. Drost, T. Ojanen, A. Harju, and P. Liljeroth, *Topological states in engineered atomic lattices*, Nat. Phys. **13**, 668 (2017).

[10] M. N. Huda, S. Kezilebieke, T. Ojanen, R. Drost, and P. Liljeroth, *Tuneable Topological Domain Wall States in Engineered Atomic Chains*, npj Quantum Mater. **5**, 17 (2020).

[11] S. N. Kempkes, M. R. Slot, J. J. van den Broeke, P. Capiod, W. A. Benalcazar, D. Vanmaekelbergh, D. Bercioux, I. Swart, and C. Morais Smith, *Robust Zero-Energy Modes in an Electronic Higher-Order Topological Insulator*, Nat. Mater. **18**, 1292 (2019).

[12] S. E. Freeney, J. J. van den Broeke, A. J. J. Harsveld van der Veen, I. Swart, and C. Morais Smith, *Edge-Dependent Topology in Kekulé Lattices*, Phys. Rev. Lett. **124**, 236404 (2020).

[13] J. A. Stroscio and D. M. Eigler, *Atomic and Molecular Manipulation with the Scanning Tunneling Microscope*, Science **254**, 1319 (1991).

[14] W. P. Su, J. R. Schrieffer, and A. J. Heeger, *Solitons in Polyacetylene*, Phys. Rev. Lett. **42**, 1698 (1979).

[15] W. P. Su, J. R. Schrieffer, and A. J. Heeger, *Soliton excitations in Polyacetylene*, Phys. Rev. B **22**, 2099 (1980).

[16] A. J. Heeger, S. Kivelson, J. R. Schrieffer, and W.-P. Su, *Solitons in Conducting Polymers*, Rev. Mod. Phys. **60**, 781 (1988).

[17] F. Stern and W. E. Howard, *Properties of semiconductor surface inversion layers in the electric quantum limit*, Phys. Rev. **163**, 816–835 (1967).

[18] S. Fölsch, J. Martínez-Blanco, J. Yang, K. Kanisawa, and S. C. Erwin, *Quantum Dots with Single-Atom Precision*, Nat. Nanotechnol. **9**, 505 (2014).

[19] S. Fölsch, J. Yang, C. Nacci, and K. Kanisawa, *Atom-By-Atom Quantum State Control in Adatom Chains on a Semiconductor*, Phys. Rev. Lett. **103**, 096104 (2009).

[20] J. Yang, C. Nacci, J. Martínez-Blanco, K. Kanisawa, and S. Fölsch, *Vertical Manipulation of Native Adatoms on the InAs(111)A Surface*, J. Phys. Condens. Matter **24**, 354008 (2012).

[21] Y. Pan, K. Kanisawa, and S. Fölsch, *Creating and Probing Quantum Dot Molecules with the Scanning Tunneling Microscope*, J. Vac. Sci. Technol. B **35**, 04F102 (2017).

[22] Y. Tong, G. Xu, and W. N. Mei, *Vacancy-Buckling Model for the (2×2) GaAs(111) Surface*, Phys. Rev. Lett. **52**, 1693 (1984).

[23] M. Morgenstern, *Probing the Local Density of States of Dilute Electron Systems in Different Dimensions*, Surf. Rev. Lett. **10**, 933 (2003).

[24] M. F. Crommie, C. P. Lutz, and D. M. Eigler, *Confinement of Electrons to Quantum Corrals on a Metal Surface*, Science **262**, 218 (1993).

[25] A. A. Khajetoorians, D. Wegner, A. F. Otte, and I. Swart, *Creating Designer Quantum States of Matter Atom-by-Atom*, Nat. Rev. **1**, 703 (2019); and references therein.

[26] J. K. Asbóth, L. Oroszlány, and A. Pályi, *A Short Course on Topological Insulators* (Springer, Heidelberg, 2016)

[27] J. A. Stroscio, R. M. Feenstra, and A. P. Fein, *Electronic Structure of the Si (111) 2×1 Surface by Scanning-Tunneling Microscopy*, Phys. Rev. Lett. **57**, 2579 (1986).

[28] See Supplementary Material at … for details.

[29] A. L. Efros, *Metal-Non-Metal Transition in Heterostructures with Thick Spacer Layers*, Solid State Commun. **70**, 253 (1989).

[30] C. Metzner, M. Hofmann, and G. H. Döhler, *Intersubband Transitions of a Quasi-Two-Dimensional Electron Gas in the Strong Disorder Regime*, Phys. Rev. B **58**, 7188 (1998).

[31] S. Perraud, K. Kanisawa, Z.-Z. Wang, and T. Fujisawa, *Imaging the Percolation of Localized States in a Multisubband Two-Dimensional Electronic System Subject to a Disorder Potential*, Phys. Rev. B **76**, 195333 (2007).

[32] M. Noguchi, K. Hirakawa, and T. Ikoma, *Intrinsic Electron Accumulation Layers on Reconstructed Clean InAs(100) Surfaces*, Phys. Rev. Lett. **66**, 2243 (1991).

[33] L. Ö. Olsson, C. B. M. Andersson, M. C. Håkansson, J. Kanski, L. Ilver, and U. O. Karlsson, *Charge Accumulation at InAs Surfaces*, Phys. Rev. Lett. **76**, 3626 (1996).

[34] K. Kanisawa, M. J. Butcher, H. Yamaguchi, and Y. Hirayama, *Imaging of Friedel Oscillation Patterns of Two-Dimensionally Accumulated Electrons at Epitaxially Grown InAs(111) Surfaces*, Phys. Rev. Lett. **86**, 3384 (2001).

[35] J. R. Weber, A. Janotti, and C. G. Van de Walle, *Intrinsic and Extrinsic Causes of Electron Accumulation Layers on InAs Surfaces*, Appl. Phys. Lett. **97**, 192106 (2010).

[36] B. Simon, *The Bound State of Weakly Coupled Schrödinger Operators in One and Two Dimensions*, Ann. Phys. (N.Y.) **97**, 279 (1976).

[37] J. Martínez-Blanco, C. Nacci, S. C. Erwin, K. Kanisawa, E. Locane, M. Thomas, F. von Oppen, P. W. Brouwer, and S. Fölsch, *Gating a Single-Molecule Transistor with Individual Atoms*, Nat. Phys. **11**, 640 (2015).

[38] P. Boross, J. K. Asbóth, G. Széchenyi, L. Oroszlány, and A. Pályi, *Poor Man's Topological Quantum Gate Based on the Su-Schrieffer-Heeger Model*, Phys. Rev. B **100**, 045414 (2019).

[39] F. Liu and K. Wakabayashi, *Novel Topological Phase with a Zero Berry Curvature*, Phys. Rev. Lett. **118**, 076803 (2017).

[40] Z. Liu, F. Liu, and Y.-S. Wu, *Exotic Electronic States in the World of Flat Bands: From Theory to Material*, Chin. Phys. B **23**, 077308 (2014); and references therein.

[41] D. Leykam, A. Andreanov, and S. Flach, *Artificial Flat Band Systems: From Lattice Models to Experiments*, Adv. Phys.: **X3**, 677 (2018); and references therein.

**Figures and legends**

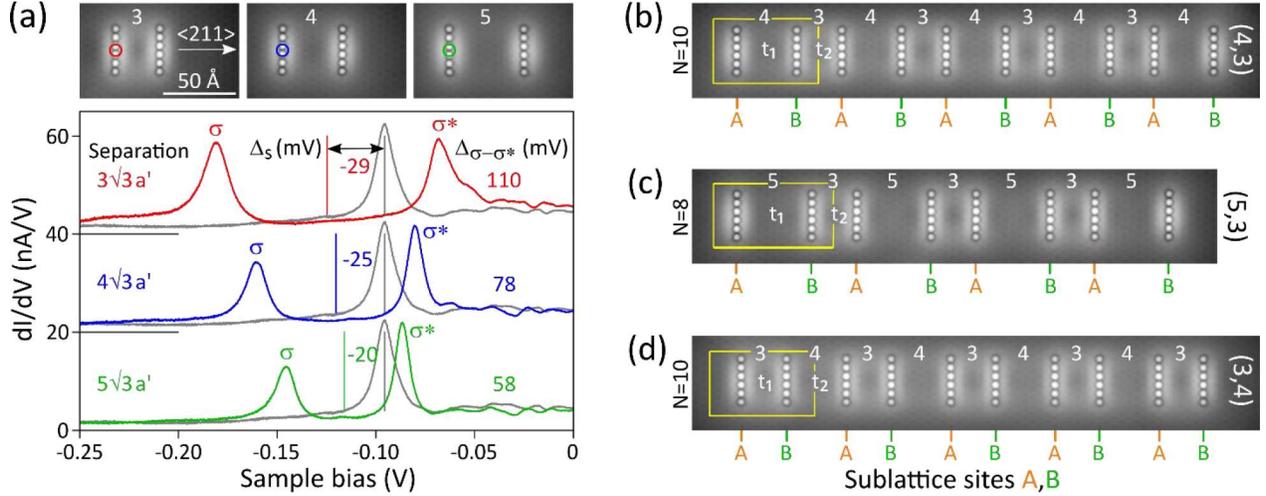

Fig. 1. (a) Upper panel: STM topography images (0.1 nA, 0.1 V) of quantum-dot dimers with interdot separations of 3, 4, and 5 in units of $\sqrt{3}a'$ (from left to right) with $a'=8.57$ Å the lattice constant of the 2×2 In vacancy reconstruction; each dot consists of six positively charged In adatoms. Lower panel: corresponding conductance spectra revealing the bonding ($\sigma$) and antibonding states ($\sigma*$) of the dimers arising from the confinement of surface-state electrons of pristine InAs(111)A, the $\sigma-\sigma*$ splitting decreases as the spacing increases (tip positions where spectra were recorded are marked by circles). The electrostatic potential experienced by each dot from the other leads to a downshift $\Delta_s$ of the $\sigma-\sigma*$ doublet relative to the corresponding confined-state energy of the discrete $In_6$ dot (gray curves). (b) STM topography image (0.1 nA, 0.1 V) of a dimer chain with alternating spacings of 4 and 3 in units of $\sqrt{3}a'$ referred to as (4,3) chains below; $N$ is the number of dots. The unit cell marked yellow contains two dots residing on $A$ and $B$ sites of the sublattice structure, respectively. The intracell hopping $t_1$ is smaller than the intercell hopping $t_2$ and the ratio is $t_1/t_2 \cong 0.7$ based on the spectra of isolated dimers in (a). (c) Same as (b) but with dot spacings of 5 and 3 [referred to as (5,3) chains], the hopping ratio is $t_1/t_2 \cong 0.5$. (d) Same as (b) but with $t_1 > t_2$ and $t_1/t_2 \cong 1.4$.

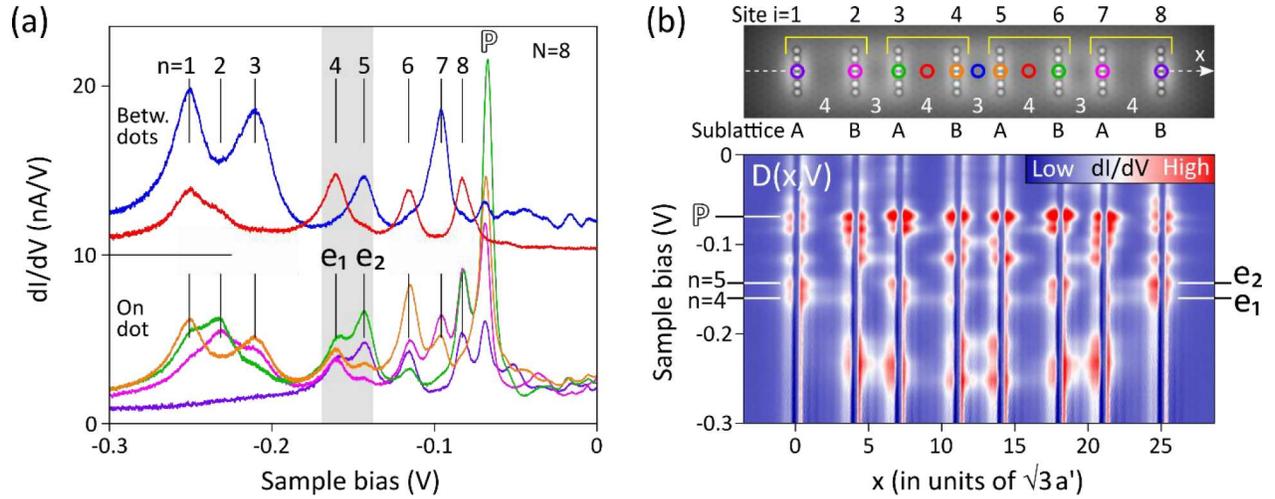

Fig. 2. (a) Conductance spectra of a (4,3) chain with $N=8$ recorded at the tip positions marked in the STM image in (b), spectra taken at symmetry-equivalent positions are averaged. The observed sequence of eight conductance peaks denoted $n=1$ to 8 reveals the eight "molecular" states arising from surface-state confinement, vertical bars are a guide to the eye. The end states $e_1$ and $e_2$ ($n=4,5$) are situated in the middle of the level spectrum (shaded gray). The extra conductance peak denoted $P$ is attributed to the electron accumulation layer near the InAs(111)A surface, cf. Fig. 4 and related discussion. (b) Upper panel: topography image (0.1 nA, 0.1 V) of the dimer chain. Lower panel: conductance map $D(x,V)$ showing the $dI/dV$ signal recorded at constant tip height along the symmetry axis of the dimer chain (dashed line denoted $x$ in the STM image) and as a function of sample bias $V$. States $e_1$ and $e_2$ have maximum probability density at the outermost dots and reside predominantly on $A$ ($B$) sites in the left (right) half of the chain.

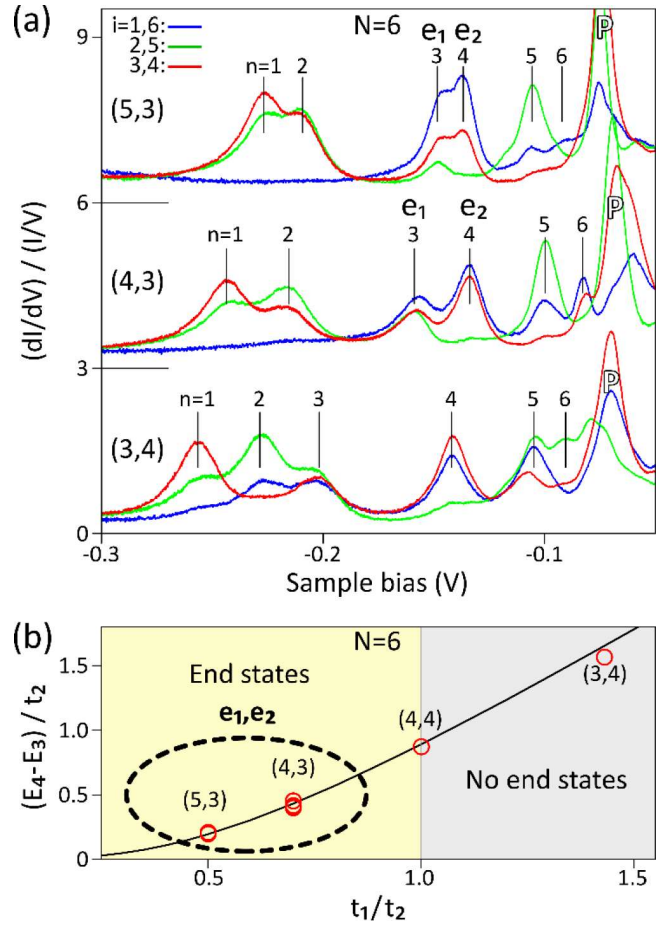

Fig. 3. (a) Normalized conductance spectra of dimer chains made of six dots in (5,3) [$t_1/t_2 \cong 0.5$, top], (4,3) [$t_1/t_2 \cong 0.7$, middle], and (3,4) configuration [$t_1/t_2 \cong 1.4$, bottom]; the latter implies alternating dot separations of 3 and 4 units with the 3-units separation at the ends. Dots are indexed $i$=1 to 6 from left to right. Spectra taken with the tip probing symmetry-equivalent dots are averaged. Vertical bars show the positions where the six molecular states are observed for each configuration. (b) Normalized level spacing $(E_4 - E_3)/t_2$ versus hopping ratio $t_1/t_2$ for dimer chains with $N$=6; energies $E_3$ and $E_4$ refer to the states with $n$=3 and $n$=4. The experimental data (red circles) are consistent with the TB result for a SSH chain indicated by the black curve. End states exist in the range where $t_1/t_2<1$ whereas they are absent when $t_1/t_2 \geq 1$.

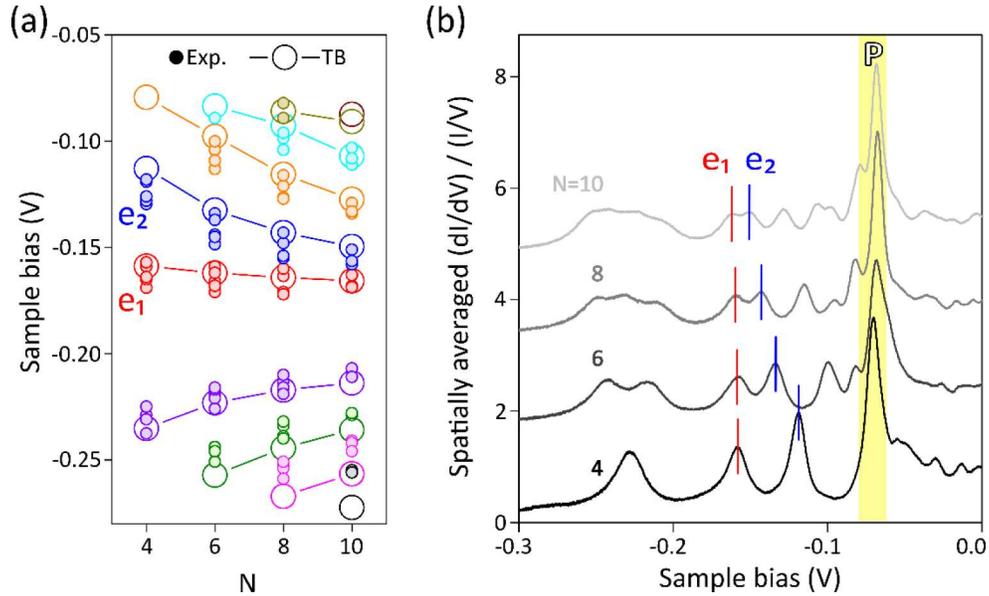

Fig. 4. (a) Experimental energies of the molecular states (small filled circles) observed for seventeen individual (4,3) dimer chains and plotted versus $N$. The scatter in the experimental data points reflects the electrostatic potential disorder due to residual charged defects in the InAs(111)A substrate. The $N$-dependent trend of the entire data set is well captured by the tight-binding (TB) calculation (large empty circles) described in the main text and Appendix B. The asymmetry in the energy level spectrum is a consequence of the varying onsite potential along the chain. (b) Normalized conductance spectra ($dI/dV$) / ($I/V$) spatially averaged over all dots of the dimer chain. While the molecular states gradually evolve and shift in energy with changing $N$, the state associated with the extra spectral peak $P$ always occurs in the same energy range (shaded yellow). This extra state is likely to arise from the confinement of the electron accumulation layer near the InAs(111)A surface.

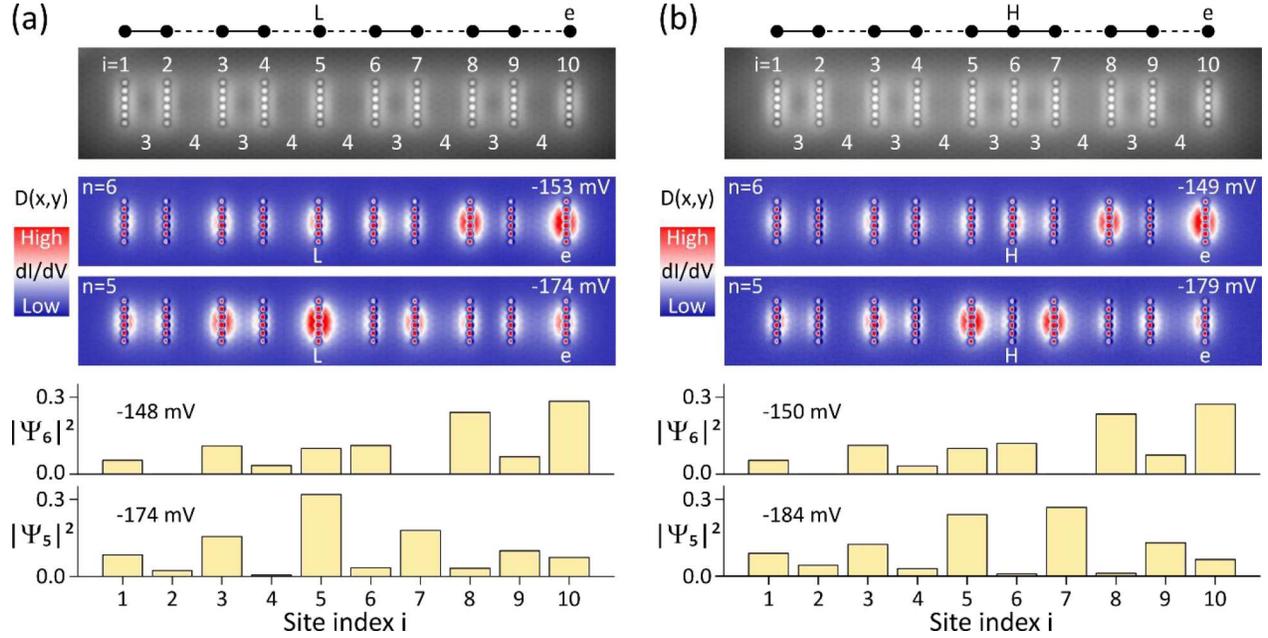

Fig. 5. (a) Upper panel: STM topography image (0.1 nA, 0.1 V) of a ten-dot dimer chain with two boundaries, a light domain wall denoted *L* and an end site of weaker bonding on the right hand-side denoted *e*. The sketch above visualizes the bonding configuration. Center panel: spatial conductance maps $D(x,y)$ of the molecular states with $n=5$ and $n=6$ reflecting the bonding and antibonding combinations of the boundary states; the state with $n=5$ ($n=6$) has enhanced weight on the domain wall state (end state). Lower panel: TB-calculated squared wave-function amplitudes $|\Psi_5|^2$ and $|\Psi_6|^2$ showing good agreement with the experimental conductance maps. (b) Same as (a) after rearranging the ten dots to form a heavy domain wall denoted *H* and an end site of weaker bonding again on the right hand-side denoted *e*. In analogy to (a), the conductance maps (center panels) and squared wave-function amplitudes reveal that the state with $n=5$ ($n=6$) has enhanced weight on the domain wall state (end state).

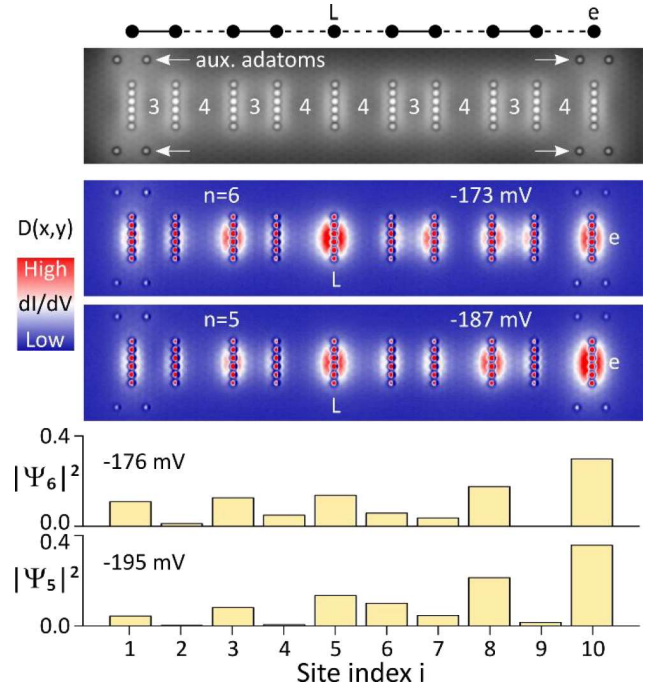

Fig. 6. Upper panel: STM topography image (0.1 nA, 0.1 V) of the ten-dot dimer chain shown in Fig. 5(a) after adding four auxiliary In adatoms (indicated by arrows) near each end dot to reduce the variation in onsite potential along the chain. The spatial conductance maps in the center panel reveal comparable probability density distributions for the even ($n$=5) and odd superposition of the boundary states ($n$=6) with a slightly larger domain-wall character at $n$=6. The TB-calculated squared wave-function amplitudes $|\Psi_5|^2$ and $|\Psi_6|^2$ in the lower panel are consistent with the experimental conductance maps.

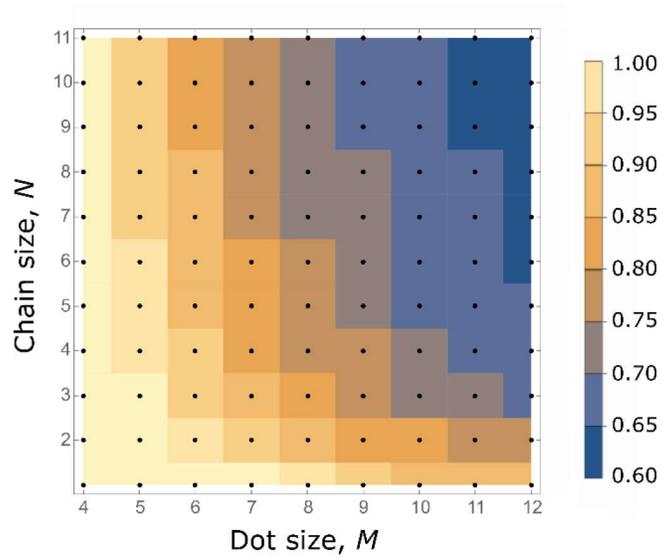

Fig. 7. Indium atomic charge $q$ that minimizes the total energy of a chain of $N$ dots each consisting of $M$ indium atoms.

# Supplementary Material to:

**Topological States in Dimerized Quantum-Dot Chains Created by Atom Manipulation**

Van Dong Pham[1], Yi Pan[1,2], Steven C. Erwin[3], Felix von Oppen[4], Kiyoshi Kanisawa[5], and Stefan Fölsch[1]

[1]*Paul-Drude-Institut für Festkörperelektronik, Hausvogteiplatz 5-7, Leibniz-Institut im Forschungsverbund Berlin e. V., 10117 Berlin, Germany*

[2] *Center for Spintronics and Quantum Systems, State Key Laboratory for Mechanical Behavior of Materials, Xi'an Jiaotong University, Xi'an 710049, China*

[3] *Center for Computational Materials Science, Naval Research Laboratory, Washington, DC 20375, USA*

[4]*Dahlem Center for Complex Quantum Systems and Fachbereich Physik, Freie Universität Berlin, 14195 Berlin, Germany*

[5]*NTT Basic Research Laboratories, NTT Corporation, 3-1 Morinosato Wakamiya, Atsugi, Kanagawa, 243-0198, Japan*

**Contents:**

## S1. Variance of absolute experimental energy values

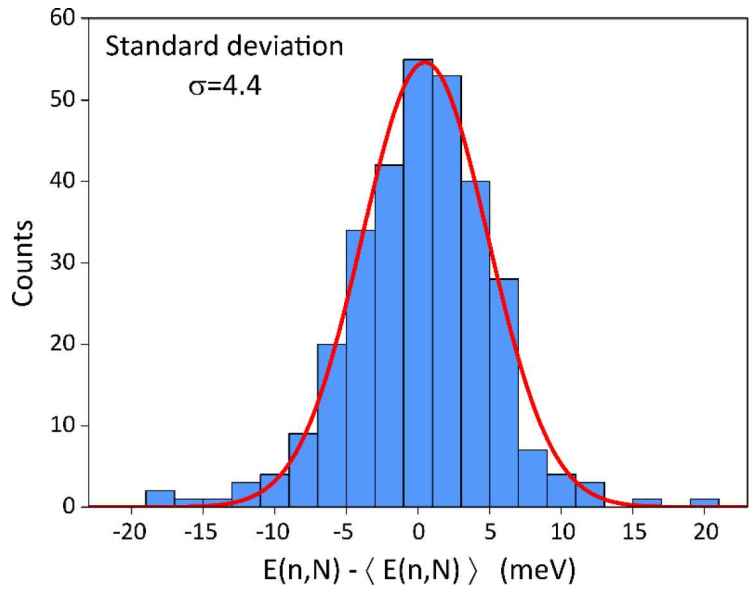

Fig. S1. Histogram of the quantity E($n,N$) − ⟨E($n,N$)⟩ where $E(n,N)$ is a single energy value and ⟨E($n,N$)⟩ the average of all single values measured at given quantum number $n$ and given number of dots $N$. The data follow a Gaussian distribution with a standard deviation σ=4.4 meV, equivalent to a full width at half maximum of ~10 meV. Dimer chains in (4,3) and (5,3) configuration as well as single dimers and quantum-dot monomers are included in the experimental data.

## S2. Chain of six evenly spaced dots

In the discussion of Fig. 3 in the main text we considered the energy levels of six evenly spaced dots implying identical nearest-neighbor hopping along the chain, $t_1=t_2\equiv t_0$. Figure S2 shows an STM topography image of this structure having an interdot spacing of 4 in units of $\sqrt{3}a'$. Also in this case, the conductance spectra reveal a sequence of molecular states plus an extra spectral feature $P$ at the high-energy side (the latter we attribute to the confinement of the surface-accumulated electrons, as discussed in the main text). A normalized energy spacing $(E_4 - E_3)/t_0 = 0.87$ is found, assuming that $t_0=\Delta_{\sigma-\sigma*}/2=39$ mV with $\Delta_{\sigma-\sigma*}$ the bonding-antibonding splitting observed for dimers separated by $4\sqrt{3}a'$ [cf. Fig. 1(a) in the main text]. This compares well with the theoretical value $-2\left(\cos\frac{4\pi}{7} - \cos\frac{3\pi}{7}\right) = 0.89$ assuming zero onsite energy.

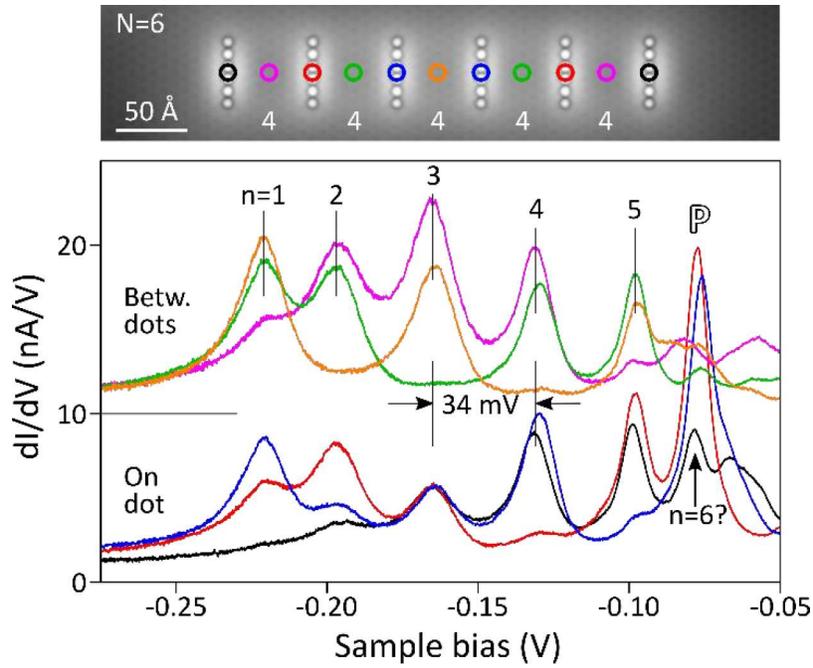

Fig. S2. Upper panel: STM topography image (0.1 nA, 0.1 V) of a chain of six equidistant dots, the separation is 4 in units of $\sqrt{3}a'$. Lower panel: Conductance spectra recorded at the tip positions shown in the STM image, spectra at symmetry-equivalent positions are averaged. Vertical bars are a guide to the eye marking the molecular states $n=1$ to 5. The highest state with $n=6$ is not unambiguously identifiable, apparently because it is obscured by the spectral feature $P$ arising from the confinement of surface-accumulated electrons.

## S3. End-state wave functions of a dimer chain with eight dots

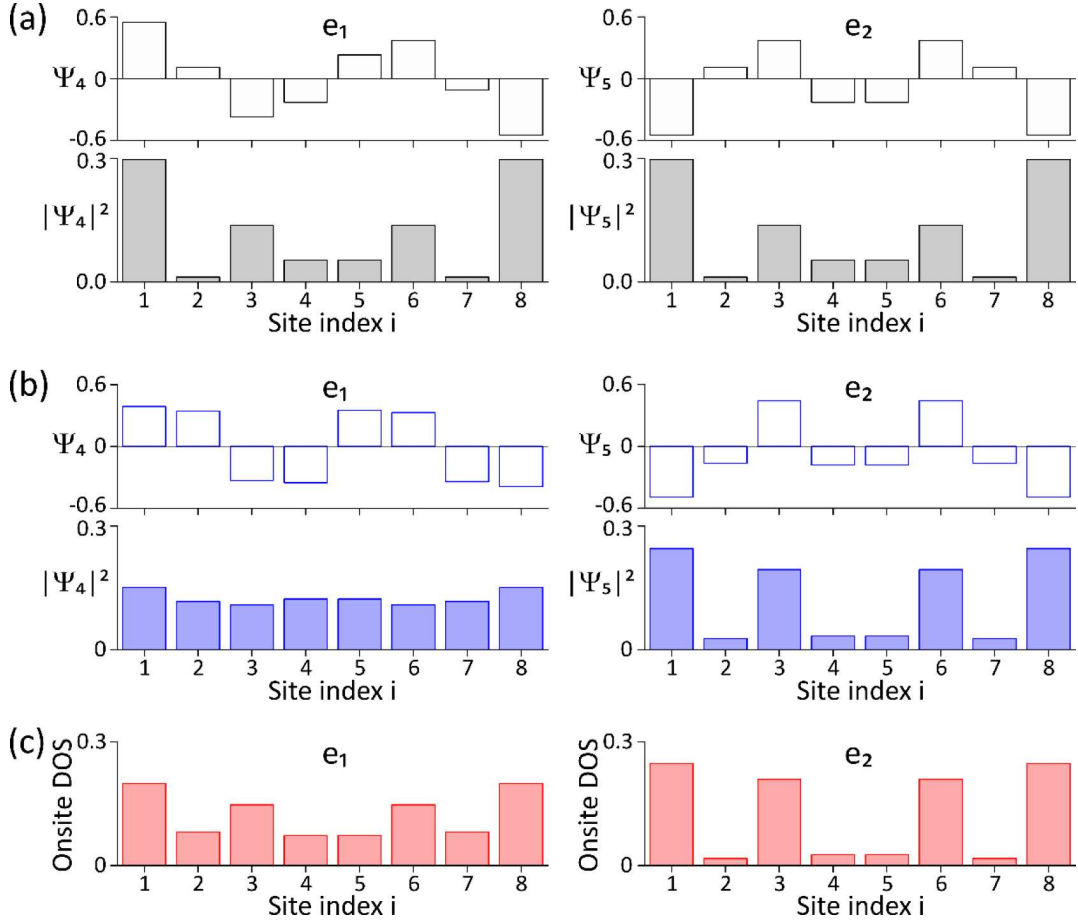

Fig. S3. (a) Tight-binding-calculated wave-function amplitude $\Psi_n$ versus site index $i$ for the odd ($e_1$, $n=4$) and even ($e_2$, $n=5$) superpositions of the (4,3) chain end states with $N=8$, hoppings $t_1$, $t_2=39$, 55 meV, and zero onsite energy for all $i$ (upper row). The wave-function envelopes resemble free-electron wave functions having $n$ lobes and $n-1$ nodes. The squared amplitudes (lower row) are identical for $e_1$ and $e_2$, which is characteristic for a finite SSH chain having a symmetric level structure. (b) Same as (a) but with varying onsite potential along the chain for charge state $q=0.69$ per adatom as obtained from a least-squares fit to the experimental energies in Fig. 4(a) of the main text. The potential creates additional confinement and hence shifts all the amplitudes toward the center of the chain. The squared amplitudes are clearly different for $e_1$ and $e_2$. (c) Normalized conductance of $e_1$ and $e_2$ measured with the tip probing the dots along a (4,3) chain with $N=8$, the sum of the obtained conductance values [providing a measure of the total density of states (DOS)] was normalized to 1 for states $e_1$ and $e_2$, respectively. Spectra taken at symmetry-equivalent positions of five independently assembled chains were averaged. The experimentally observed asymmetry between $e_1$ and $e_2$ is well reproduced by the TB result for varying onsite potential in (b).